\documentstyle[12pt,epsfig]{article}
\begin{document}
\baselineskip=2pc
\def\bfg #1{{\mbox{\boldmath $#1$}}}
\begin{center}
{\large \bf { BACKWARD ELASTIC $p~^3He$-SCATTERING
AND HIGH MOMENTUM COMPONENTS OF $~^3He$ WAVE FUNCTION
   }}
\vskip 1em
 Yu.N. Uzikov

{\it Laboratory of Nuclear Problems,
 Joint Institute for Nuclear Research,\\
 Dubna, Moscow reg., 141980 Russia}
\end{center}

\section*{ Abstract}
 It is shown that owing to a dominance of $np-$pair transfer mechanism of
 backward elastic $p~^3He$- scattering  for incident  proton kinetic
  energies  $T_p>1$ GeV the cross section of this process is  defined
 mainly by the values  of the Faddeev component of the wave function of
 $~^3He$ nucleus, $\varphi ^{23}({\bf q}_{23}, {\bf p}_1)$, at high
 relative momenta  $q_{23}> 0.6 GeV/c$ of the NN-pair in the $^1S_0-$ state
 and at low spectator momenta $p_1\sim 0-0.2$ GeV/c.

PACS numbers: 25.10.+s, 25.40.Cm, 21.45.+v
\vskip 1.5em

  The cross section of backward elastic $p~^3He$-scattering at
 the kinetic energy  of incident proton $T_p> 1$ GeV displays three
 remarkable peculiarities \cite{laduz92,[2]}. (i) In the Born approximation
 only one mechanism of the process $p^3He\to ~^3Hep$ dominates, it is the
 so-called sequential  transfer (ST) of the noninteracting np-pair.
 The contribution from the  mechanisms of nonsequentional transfer (NST),
 interacting np-pair  transfer (IPT) and deuteron exchange  is negligible.
 The heavy particle stripping mechanism was also investigated in Refs.
  \cite{gurvitz} -\cite{sherif} and found to be important at back angles
  for $T_p\leq 0.6$ GeV. However the phenomenological $^3He$ wave functions
  restricted to  the two-body configuration, which does not permit of
  ST-mechanism  were used in that analysis.
 (ii) The  most  important role in the Faddeev wave function
 $\varphi^{23}({\bf q}_{23}, {\bf p}_1)$  of $^3He$ plays the channel with
 the orbital momentum $L=0$, spin $S=0$, isotopic spin $T=1$ of two nucleons
 with numbers 2 and 3 and the orbital momentum $l=0$ of nucleon spectator
 with number 1 (it coresponds to $\nu=1$ in the notation of Ref.\cite{bkt}).
 If this channel is excluded from the full wave function
 $\Psi=\varphi^{23}+ \varphi^{31}+\varphi^{12}$, the cross section falls
 by several orders of magnitude. (iii) Rescatterings in the initial and final
 states decrease the cross section
 at  $\theta_{c.m.}=180^0$  considerably in comparison with the Born
 approximation and make it agree satisfactorily with
 the available experimental data \cite{berth} for $T_p> 0.9$ GeV.

 Owing to this evident connection between the structure of $~^3He$
 nucleus and the dominating mechanism one can hope to obtain an information
 about high momentum components of the $~^3He$ wave function from the cross
 section  of the $p^3He\to ~^3Hep$ process. However, in Refs.
 \cite{laduz92,[2]} it was mentioned that the D-components of $^3He$ wave
 function  are of surprisingly
 minor importance in the process under discussion at $T_p> 1$ GeV. Moreover,
 relativistic effects estimated in Ref. \cite{[2]} at $T_p\sim 1$ GeV
 by means of substituting the relativistic argumets into the $~^3He$ wave
 function instead of the nonrelativistic ones give rather small contribution
 into the cross section.
 For this reason, in Refs. \cite{laduz92,[2]} it was  concluded  that
 the sensitivity of the  $p~^3He\to ~^3Hep$ cross section to the high
 momentum components of the $~^3He$ wave function is rather weak in spite of
 high  momenta transferred at $T_p>1$ GeV. Moreover, as was found in
 \cite{berth},
 the role of the triangular diagram of one-pion exchange (OPE) with the
 subprocess  $pd\to ~^3He\pi^o$ related to the $\Delta$ - and double
 $\Delta$-excitation is in qualitative agreement
  with the absolute value of the experimental cross section at
$T_p> 0.5$ GeV.

  In the present  work it is shown  that  the absolute value
 of the  $p~^3He\to ~^3Hep$ cross section at  $\theta_{c.m.}=180^o$
 and $T_p>1$ GeV  is directly related to the high momentum components
 of the Faddeev S-wave function of $^3He$, $\varphi^{23}({\bf q}_{23},
 {\bf p}_1)$, respecting the relative momentum ${\bf q}_{23}$ whereas
  rather low  values of  the  "spectator" momentum ${\bf p}_1$
 are involved into the amplitude of this process. It is shown also, that
 due to rescatterings in the initial and final states the contribution of
 the OPE mechanism is  one order of magnitude lower in comparison with
 the experimental data.

 In the Born approximation the amplitude of  transfer of two nucleons
   with numbers 2 and 3 in the process $0+\{123\}\to 1+\{023\}$
 (et id. $p~^3He\to ~^3Hep$) can be written as  \cite{laduz92,[2]}
$$T_B=6(2\pi )^{-3}\int d^3q_{23}L_{23}(q_{23},p_1)
\chi^+_{p'}(1)\bigl \{\varphi^{{23}^+}_f(0;23) \varphi^{31}_i(2;31)+$$
 \begin{equation}
\label{u1}
\varphi_f^{{02}^+}(3;02) \varphi_i^{31}(2;31)+
\varphi_f^{{30}^+}(2;30) \varphi_i^{31}(2;31)\bigr \}\chi_{p}(0),
\end{equation}
 where $\varphi^{ij}(k;ij)=\varphi^{ij}({\bf q}_{ij},{\bf p}_k)$
 is the Faddeev component of the wave function of the bound
 state $\{ijk\}$, $\chi_{p}(\chi_{p'})$ is the spin-isotopic spin
 wave function of the incident (final) proton;
 $L_{23}=\varepsilon+{\bf q}^2_{23}/m+3{\bf p}^2_1/4m$, $m$ is the nucleon
 mass, $\varepsilon $ is the $~^3He$ binding energy.
 The subscripts {\it i} and {\it f} in Eq.( \ref{u1}) refer to the initial
 and final nucleus respectively.
 The terms $\varphi^{{23}^+}_f \varphi^{31}_i $,
 $\varphi^{{02}^+}_f \varphi^{31}_i$,
 $\varphi^{{30}^+}_f \varphi^{31}_i$ correspond  to the IPT, ST and
 NST mechanisms respectively.   In the explicit form the ST mechanism
 has the following structure of arguments
$$\varphi^{{02}^+}_f \varphi^{31}_i=
 \varphi^{{02}^+}_f
\bigl ({\bf q}_{02}=-\frac{1}{2}{\bf q}_{23}-\frac{3}{4}{\bf Q}_0,
{\bf p}_3={\bf q}_{23}-\frac{1}{2}{\bf Q}_0 \bigr )\,$$
\begin{equation}
\label{u2}
 \times\varphi^{{31}}_i
\bigl ({\bf q}_{31}=-\frac{1}{2}{\bf q}_{23}+\frac{3}{4}{\bf Q}_1,
 {\bf p}_2=-{\bf q}_{23}-\frac{1}{2}{\bf Q}_1 \bigr ),
 \end{equation}
 where ${\bf Q}_0\, ({\bf Q}_1)$ is the momentum of incident (final) proton
 in the c.m.s of the final (initial) nucleus $~^3He$. As was noted in
\cite{[2]}, at the scattering angle $\theta_{c.m.}=180^o$ two of four momenta
 in Eq.(\ref{u2}) can simultaneously become equal to zero at integration over
 ${\bf q}_{23}$. On the contrary, in the corresponding formulas for the
 IPT and NST mechanisms only one argument can be equal to zero while
 the other three have large values $\sim |{\bf Q}_1|= |{\bf Q}_0|$.
 This makes the ST-term dominate  in Eq.(\ref{u1}).
 Indeed, the ST-mechanism takes place only if the channels with the isotopic
 spin $T=1$ of the pair of nucleons \{ij\} are included into the component
 $\varphi^{ij}(ij;k)$ either in the initial or final state.
 It is the direct consequence of the
 fact that the ST diagram either starts with or ends in the pp-interaction.
 The $~^3He$ wave function  from Ref. \cite{bkt} contains only one
 such channel ($\nu=1$), namely, with the $^1S_0$ state of the NN-pair.
 In the S-wave approximation for the $~^3He$ wave function the cross
 section decreases by  5-6 orders of magnitude for $T_p>1$ GeV if the channel
 $\nu=1$ is excluded \cite{luz93}. The channels with $\nu \not = 1$
 coresponding to the isotopic spin $T=0$ of the NN-pair (in particularly,
 the D-components) can enter the ST-amplitude only in
 combination with the channel $\nu=1$. For this reason the role of those
 channels is not so significant.

    An obvious modification of formalism of the triangular OPE diagram from
 Refs. \cite{zhuz81,naksat} is used here for the OPE amplitude. The cross
 section of the $p~^3He\to ~^3Hep$ process is expressed through the cross
 section  of the reaction $pd\to ~^3He\pi^0$, which is taken here  from
 the experimental data \cite{berth85}. The overlap integral of $~^3He$ and
 deuteron wave functions, $<~^3He|d,p>$, is taken from \cite{konuz97}.
 Rescatterings in the initial and final states  for the  OPE mechanism are
 taken here in the line of  work \cite{[2]} on the basis of Glauber-Sitenko
 approximation.

\begin{figure}[hbt]
\mbox{\epsfig{figure=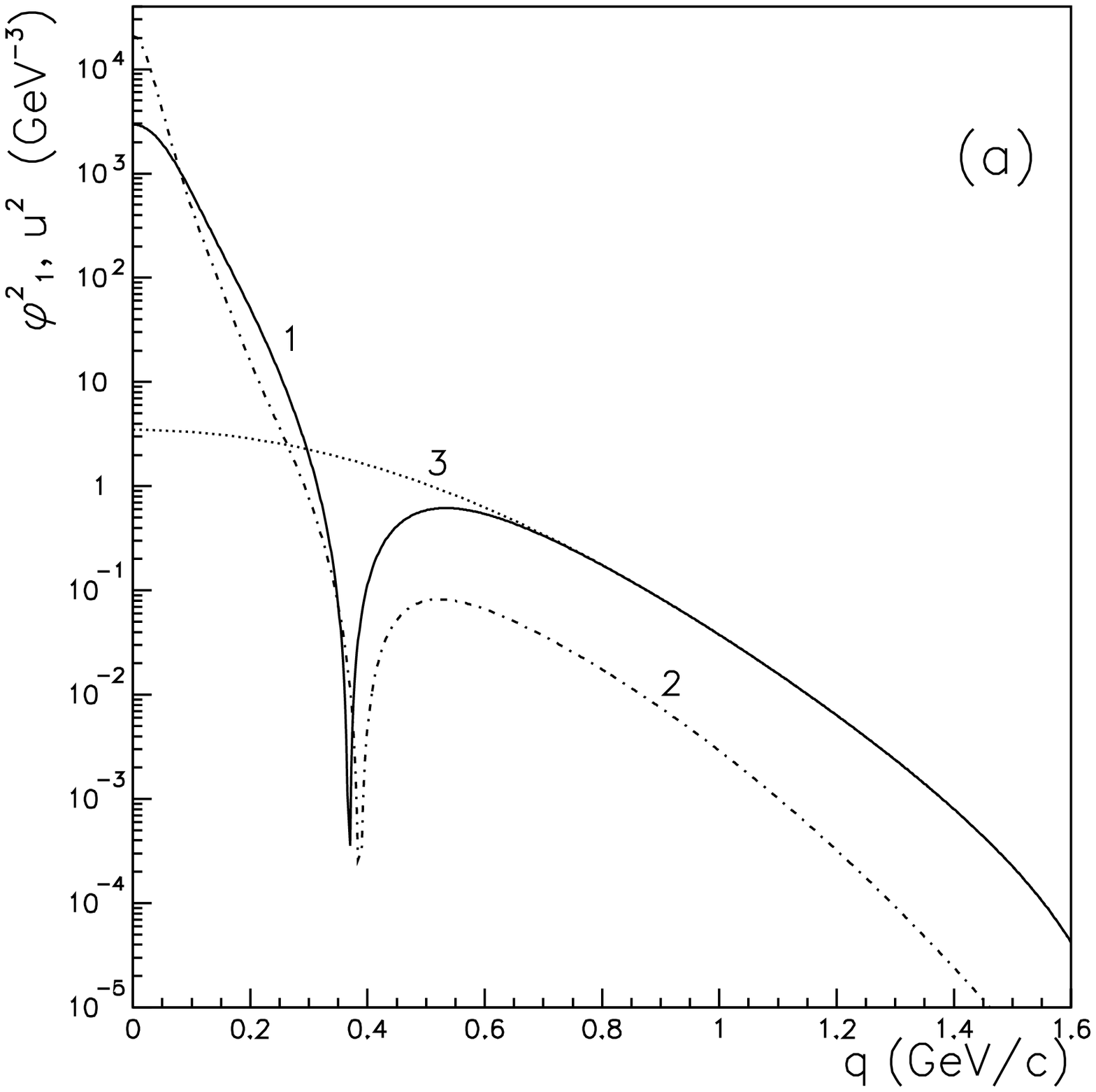,height=0.70\textheight, clip=}}
\caption{}
\label{fig1}
\end{figure}

  Numerical calculations for the $np$-pair transfer mechanism are
  performed here on the basis of the formalism
  described in \cite{laduz92,[2]} using the  $~^3He$
 wave function obtained in Ref. \cite{bkt} from the solution of Faddeev
 equations in momentum space for the RSC interaction potential between
 nucleons in the $^1S_0$ and  $^3S_1-^3D_1$ states.
 The separable  analytical parametrization for the $~^3He$ wave function
  is used here which has the  following form in terms of the notation
   \cite{hgs}
 \begin{equation}
 \label{u3}
 \phi_\nu=n_\nu \varphi_\nu(q_{23})\chi_\nu(p_1).
 \end{equation}
 The square of the functions $ \varphi_\nu(q), \chi_\nu(q)$ and
 the S-component of the deuteron wave
 function, $u(q)$, for the RSC potential \cite{alberi} are shown  in
 Fig.\ref{fig1}.
  The calculated  differential cross section is shown in Fig.\ref{fig3}
 in comparison with the experimental  data  \cite{berth}.

 The  numerical results demonstrate the following important features of
 the process in question.  Firstly,  the ST-mechanism involves the high
 momentum components  of the functions $\varphi_\nu(q_{23})$ for the
 S-wave states.  The $^3He$ wave function in the
 channel $\nu=1$ is probed at  high momenta ${\bf q}_{23}> 0.6 GeV$
 when the cross section is measured at $T_p> 1$ GeV. To show it,
 in Fig.\ref{fig1} ({\it a}) we present a part  of the function
 $\varphi_1(q_{23})$, denoted as ${\widetilde \varphi}_1$, which coincides
 with  $\varphi_1(q_{23})$
 for $q_{23}>0.6 GeV/c$ and considerably  differs from it for smaller momenta
 $q_{23}<0.5 GeV/c$. In Fig.\ref{fig1} ({\it b}) we also show a  part of the
 function $\chi_1 (p_1)$, denoted as ${\widetilde \chi}_1$,
 which is very close to the total function $\chi_1 (p_1)$ at small spectator
 momenta $p_1\sim 0-0.2 $GeV/c and is negligible for $p_1>0.2 $GeV/c.
 The cross section calculated with these two parts instead of the
 full functions  $\varphi_1$ and $\chi_1 $ is shown in Fig.\ref{fig3} by
 curves 2 and 3  respectively.  One can see that these curves are  very close
 to  the total result obtained with the full functions  $\varphi_1(q_{23})$
 and  $\chi_1 (p_1)$. In contrast, it can be shown that the cross section
 calculated with the  complementary  parts $\varphi_1-{\widetilde \varphi}_1$
 and $\chi_1-{\widetilde \chi}_1$ is 5-6 orders of magnitude smaller.

  Secondly, the above  result  displays also that the ST-mechanism
  involves rather low  "spectator"-momenta $p_1\sim 0-0.2GeV/c$ in the
 function $\chi_\nu(p_1)$ , which makes this mechanism
 dominate.  The qualitative explanation of these results is following.
 One can find from Eq.(\ref{u2}),  that for ${\bf Q}_1=-{\bf Q}_0$ (et id.
 $\theta_{c.m.}=180^o$) the equations ${\bf q}_{31}={\bf q}_{02}$ and
 ${\bf p}_{2}=-{\bf p}_{3}$ are satisfied. Consequently, the main
 contribution into the  integral over $d{\bf q}_{23}$ in Eq. (\ref{u1})
 gives the region $|{\bf p}_{2}|=|{\bf p}_{3}|\sim 0$, in which
 $|{\bf q}_{31}|=|{\bf q}_{02}| \sim Q_1$. On the contrary,
 the region of $|{\bf q}_{31}|=|{\bf q}_{02}| \sim 0$ corresponds to
 $|{\bf p}_{2}|=|{\bf p}_{3}|\sim 2Q_1$ and plays insignificant role since
 for $T_p>1$ GeV the momentum $Q_1$ is large, $Q_1> 0.6$ GeV.
\footnote{ The replacement of  the nonrelativistic momenta
 $Q^{nr}_1=Q^{nr}_0$ by
 the relativistic ones $Q^{rel}_1=Q^{rel}_0$ (where $Q^{rel} < Q^{nr}$)
 practically does not change  the ST-cross section at energies
 $T_p=0.4-1.2 $ GeV as was found in \cite{[2]}. However, for
 energies $T_p>1$ GeV such replacement
 becomes important  and increases the cross section. Therefore, in complete
 future analysis of this process one  should take into account
 relativistic effects in a consistent way.}

 Thirdly, we have find numerically, that the contribution
 of the OPE mechanism without taking into account rescatterings is in
 agreement with the experimental data at $T_p=0.5-1.3$ GeV, but is by factor
 $\sim 20-30$ smaller in comparison with the ST-contribution in the Born
 approximation for $T_p>0.8$ GeV. After allowance for rescatterings in
 the initial and final states the contribution of the OPE mechanism decreases
 by one order of magnitude and becames considerably  lower then the
 experimental data (Fig. \ref{fig3}). Probably, the cross section
 of the $p~^3He\to ~^3Hep$ process for $T_p< 1 GeV$ is defined mainly by
 the multistep $pN$-scattering mechanisms discussed in Refs.\cite{pl,ls}
 and heavy-particle stripping mechanism \cite{gurvitz} -\cite{sherif} also.
 We stress that the high momentum components of the functions $\varphi_\nu$
 in Eq.(\ref{u3}) play the most important role in the competition between the
 OPE and ST mechanisms. One can see from Fig.1 ({\it a}), that  the high
 momentum component in the function  $\varphi_1(q)$ is  richer in comparison
 with the deuteron wave  function  $u(q)$, especially for $q> 0.5$ GeV/c.
 Actually, when  substituting the S-component of the deuteron wave function
 $u(q)$  into Eq.(\ref{u3})  instead of the function $\varphi_\nu(q_{23})$
 for  $\nu=1$ and $2$ one finds  the np-transfer cross section
 decreasing by a factor $\sim $40 (curve 4 in Fig. \ref{fig3})
 and  becoming comparable  in absolute value with the OPE contribution
 in the Born approximation. Note in this connection that in $pd\to dp$
 process the contribution of the neutron exchange mechanism in the Born
 approximaion is not dominating \cite{azh} for $T_p> 1$GeV and  comparable
 with the OPE mechanism \cite{naksat, uz979}.

  As an additional test of the np-transfer mechanism the spin-spin
 correlation parameter $\Sigma $ is calculated here  for the process
 with  a polarized incident proton and a nucleus.
 This parameter is defined as
\begin{equation}
\label{u5}
\Sigma=\frac{d\sigma(\uparrow \uparrow)/d\Omega-d\sigma(\uparrow \downarrow)/
 d\Omega}
 {d\sigma(\uparrow \uparrow)/d\Omega+d\sigma(\uparrow \downarrow)/d\Omega},
\end{equation}
where $d\sigma(\uparrow \uparrow)/d\Omega$ and
$d\sigma(\uparrow \downarrow)/d\Omega$ are
the cross sections for  parallel and antiparallel spins of colliding
 particles respectively. The numerical calculations with alowance for two
 channels $\nu=1$ and $\nu=2$ in the  $~^3He$  wave function show that
at $\theta_{c.m.}=180^o$ and $T_p\sim 1-2.5$ GeV the value $\Sigma$ is
$\sim 0.1 -0.15$ independently of the initial energy.

 In conclusion, the remarkable sensitivity of the cross section of backward
 elastic $p~^3He$-scattering to the high momentum components of the
 $~^3He$ wave function in the S-wave channel is found for energies above 1
  GeV.  The  dominance of nucleon degrees of freedom is demonstrated.
 Since the mechanism of the np-pair transfer describes  the available
 experimental data in the interval of incident energies 0.9-1.7 GeV
 satisfactorily, there is a reason to measure the cross section
 at higher energies in order to enlighten the validity  of phenomenological
 NN-potentials  in describing the structure of lightest nuclei at high
 relative  momenta of nucleons.

 This work was supported in part by the Russian Foundaion for Basic Research
 (grant $N^o$ 96-02-17215).
\eject
\section*{Figure captions}
 {\bf Fig.1} The square of functions
 $\varphi _1 (q)$,  $\chi_1(q)$ from Ref. \cite{hgs},
  the S-component of the deuteron wave function  $u(q)$
from Ref. \cite{alberi} and functions $\widetilde  {\varphi}_1(q)$
 and ${\widetilde{\chi}}_1(q)$
 defined in the text. ({\it a}) 1 -- $\varphi^2 _1 (q)$,
2 -- $u^2(q)$, 3 -- ${\widetilde{\varphi}}_1^2(q)$; ({\it b})
 1 -- $\chi^2_1(q)$, 2 --  ${\widetilde{\chi}}_1^2(q)$.

 {\bf Fig.2} The differential cross section of $p^3He$ elastic scattering
 at $\theta_{c.m.}=180^o$ as a function of incident proton kinetic
 energy $T_p$. Curves 1-4 show the results of calculation in the
Born approximation for  amplitude in Eq. (\protect\ref{u1}):
 1 --   with the $~^3He$ wave function from \protect\cite{hgs},
 2 -- with ${\widetilde {\varphi}}_1(q_{23})$  instead ${\varphi}_1(q_{23})$,
 3 -- with  ${\widetilde{\chi}}_1(p_1)$ instead of ${\chi}_1(p_1)$,
 4 --  with the deuteron w.f. $u(q_{23})$ instead of
 ${\varphi}_1(q_{23})$ and ${\varphi}_2(q_{23})$.
 The results obtained with allowance for rescatterings in the initial
and final states are shown by curves 5 and 6:
5 -- the $np$-transfer mechanism
 with the $~^3He$ wave function from \protect\cite{hgs}, 6 -- OPE.
 The experimental points are taken from Ref.\protect\cite{berth}
\eject

\eject
\eject
\begin{figure}[hbt]
\mbox{\epsfig{figure=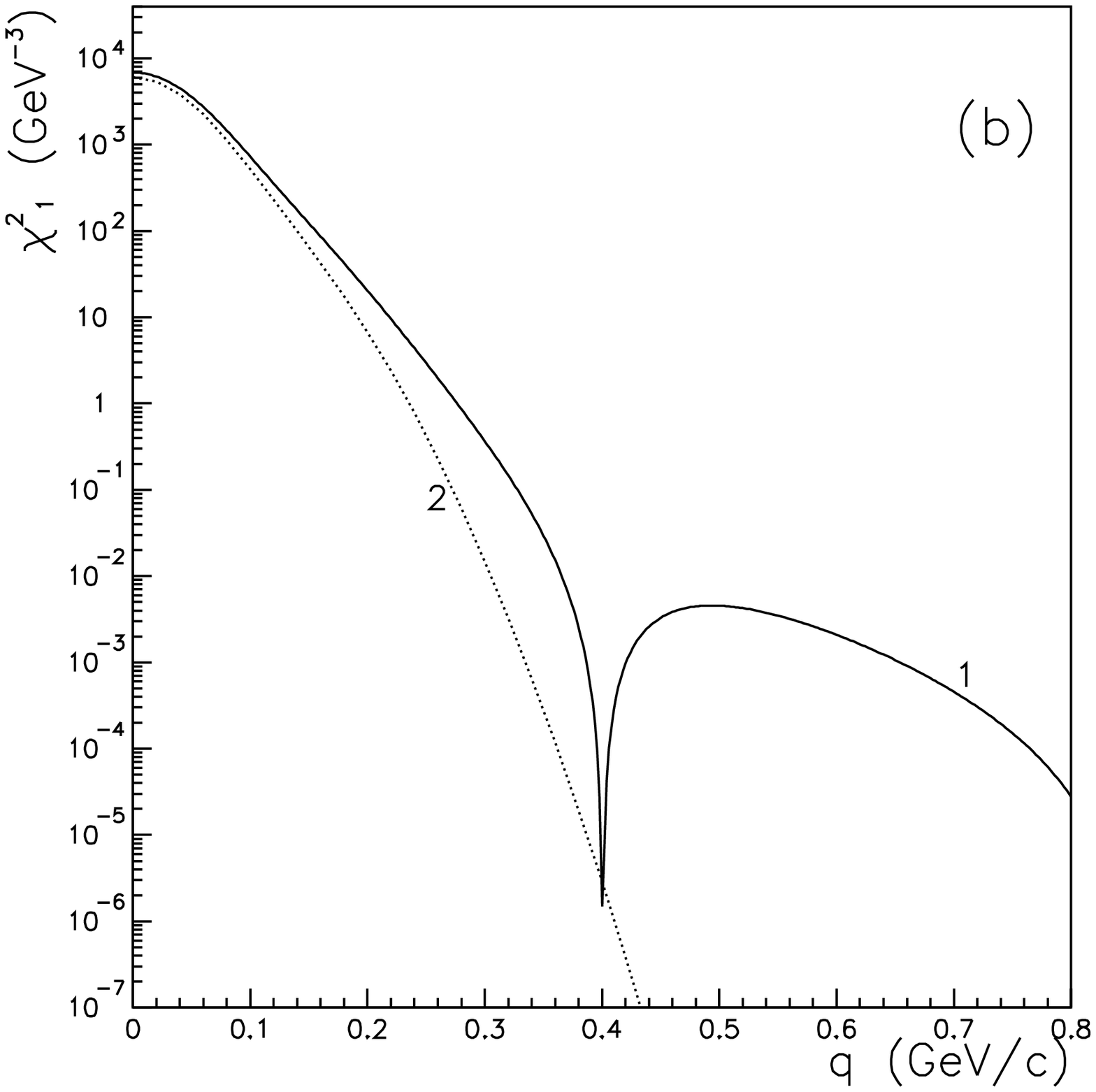,height=0.7\textheight, clip=}}
%\caption{}
\label{fig2}
\end{figure}
\eject
\begin{figure}[hbt]
\mbox{\epsfig{figure=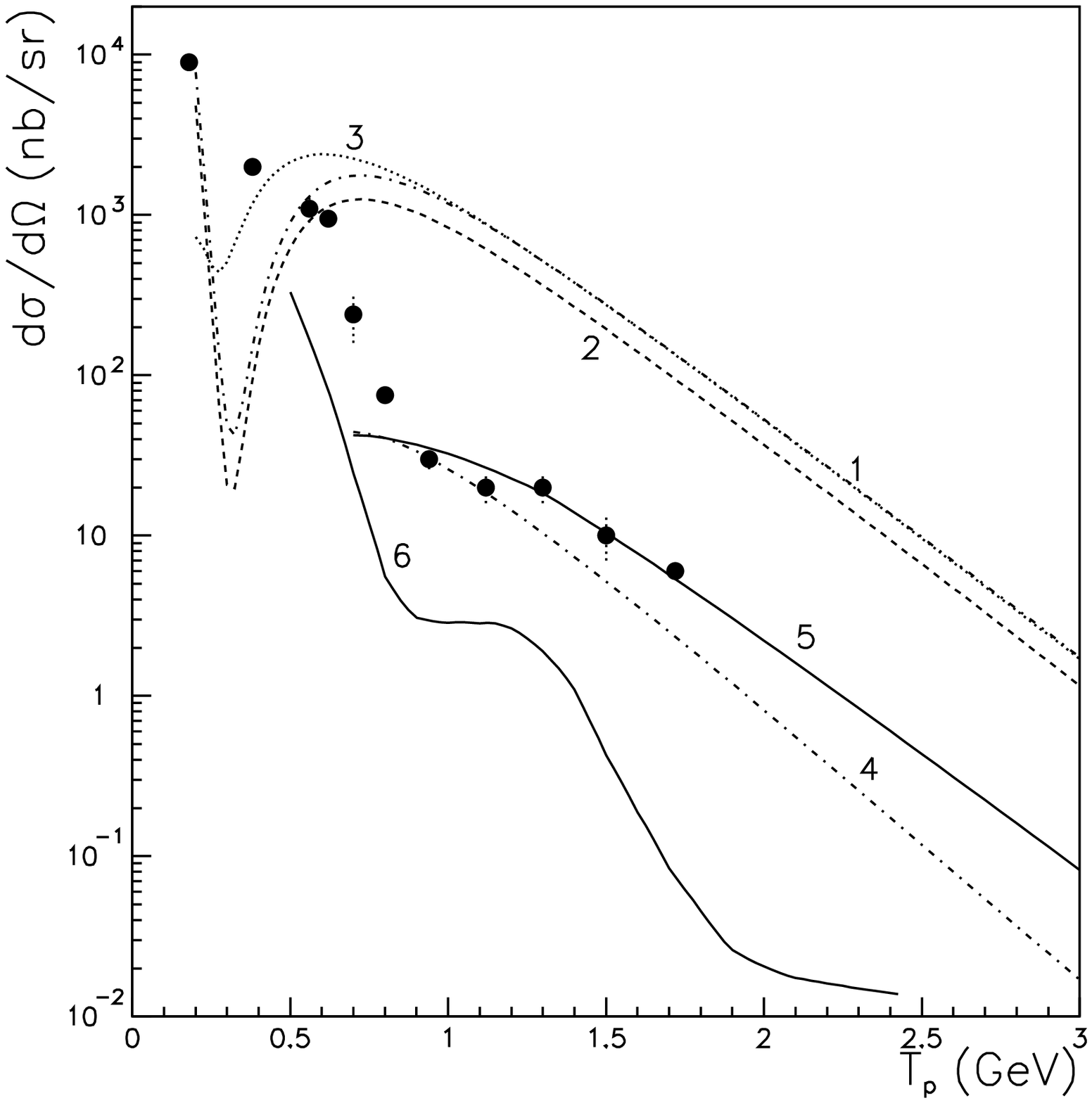,height=0.7\textheight, clip=}}
\caption{}
\label{fig3}
\end{figure}
\end{document}